\documentclass[twocolumn,secnumarabic,amssymb, amsmath, nofootinbib,tightenlines,
nobibnotes, aps, prl,epsfig]{revtex4}
\usepackage{graphicx}
\usepackage{dcolumn}
\usepackage{bm}
\begin{document}
\preprint{APS/123-QED}
\title{Effect of the parameterization of the distribution functions on the  longitudinal structure function at small $x$}

\author{G.R.Boroun}%
 \email{grboroun@gmail.com; boroun@razi.ac.ir }
\affiliation{ Physics Department, Razi University, Kermanshah
67149, Iran}
\date{\today}
\begin{abstract}
I use a direct method to extract the longitudinal structure
function in the next-to-leading order approximation with respect
to the number of active flavor from the parametrization of parton
distributions. The contribution of charm and bottom quarks
corresponding to the gluon distributions (i.e.,
$G_{n_{f}=3}(x,Q^{2})$ and $G_{n_{f}=5}(x,Q^{2})$) is considered.
I compare the obtained longitudinal structure function at
$n_{f}=4$ with the H1 data [Eur.Phys.J.C{\bf74}, 2814(2014) and
Eur.Phys.J.C{\bf71}, 1579 (2011)] and the result L.P.Kaptari et
al.[ Phys.Rev.D{\bf 99}, 096019(2019)] which is based on the
Mellin transforms. These calculations compared with the results
 from CT18 [Phys.Rev.D103, 014013(2021)] parametrization model. The nonlinear effects on the gluon distribution
 improve the behavior of the longitudinal structure function in comparison with the H1 data and CT18 at low values of $Q^{2}$. \\

\end{abstract}
 \pacs{***}
\keywords{****} 
\maketitle
In the past 20 years, the HERA experiments (i.e., H1 and ZEUS)
[1-3] have extended the knowledge of the longitudinal proton
structure function. First measurements of the longitudinal proton
structure at small $x$ were performed at HERA, as the data
collected for the longitudinal structure function were taken with
lepton beam energy of $27.6 ~\mathrm{GeV}$ and proton beam
energies $E_{p}$ of 920, 575 and $460~\mathrm{GeV}$. The HERA
measurements are covered the regions of $3{\times}10^{-5}<x<0.03$
and $1.5{\leq}Q^{2}{\leq}800~\mathrm{GeV}^{2}$ with maximum
inelasticity $y=0.85$. In the future the ep colliders (i.e., the
LHeC  and the FCC-eh) will be generated and extended to much lower
values of $x$
and high $Q^{2}$ [4].\\
It was shown that power-like corrections to $F_{L}$ is essential
more important then  power-like corrections to $F_{2}$ [5]. The
$F_{L}$ measurement is directly sensitive to the gluon
distribution in the proton [6,7]. The contribution of $F_{L}$ to
the reduced cross section is significant only at high $y$ as this
behavior has been predicted by Altarelli and Martinelli  [8] when
including higher order QCD terms. Indeed this represents a crucial
test on the validity of the perturbative QCD (pQCD) framework at
small $x$, when compared the experimental data with theoretical
predictions. To predict the rates of the various processes a set
of  gluon distribution functions corresponding to the active
flavor number $n_{f}$ is required. When $Q^{2}$ increases above
$m_{c}^{2}$ and then above $m_{b}^{2}$, the number of active
flavors increases from $n_{f}=3$ to $n_{f}=4$ and then to
$n_{f}=5$, which corresponds to the variable-flavor-number scheme
(VFNS) [9]. The charm and bottom quarks are considered as
infinitely massive below $Q^{2}=m^{2}_{c,b}$ and massless above
this threshold. For realistic kinematics it has to be extended to
the case of a general- mass VFNS (GM-VFNS) which is defined
similarly to the zero-mass VFNS (ZM-VFNS) in the
$Q^{2}/m^{2}_{c,b}{\rightarrow}\infty$ limit. For scales
$Q^{2}<m^{2}_{c}$ the fixed- flavor- number scheme (FFNS) is valid
and for $Q^{2}>m^{2}_{c}$ the approach  outlined above to define a
VFNS is valid [10].\\
In the present paper  the behavior of the longitudinal structure
function $F_{L}(x,Q^{2})$ at small $x$ by employing the
parametrization of $F_{2}(x,Q^{2})$ and $G_{n_{f}}(x,Q^{2})$
presented in Refs. [11,12] investigated. I demonstrate that, the
small $x$ behavior of the longitudinal structure function can be
directly related to the gluon behavior due to the number of active
flavors $n_{f}$ [12]. In the first step of this analysis, the
method applies to extract $F_{L}(x,Q^{2})$ in the leading-order
(LO) approximation. Then, in the region of ultra-low values of
$x$, the next-to-leading order (NLO) corrections become
significant and are to be implemented in to the extraction
procedure. Similar investigations of the longitudinal structure
function have been performed at LO and NLO approximation in Refs.
[13-29]. I provide further development of the method extending by
considering and resumming the NLO corrections for light and heavy
quarks production.\\
In pQCD, the Altarelli-Martinelli (AM) equation for longitudinal
structure function in terms of the number of active flavor at
small $x$ can be written as
\begin{eqnarray}
x^{-1}F_{L}=<e^{2}>(C_{L,q}\otimes q_{s}+C_{L,g}\otimes g),
\end{eqnarray}
where $<e^{2}>$ stand for the average of the charge $e^{2}$ for
the active quark flavours
($<e^{2}>=n^{-1}_{f}\sum_{i=1}^{n_{f}}e_{i}^{2}$).   Here
$q_{s}(x,Q^{2})$ and $g_{n_{f}}(x,Q^{2})$ are the flavour singlet
and gluon density ($G=xg$).   The $\otimes$ symbol denotes the
convolution integral which turns into a simple multiplication in
Mellin $N$-space and the notation is given by $a(x)\otimes
b(x)=\int_{x}^{1}\frac{dz}{z}a(z)b(\frac{x}{z})$. The coefficient
functions can be expressed as $C_{L,q \&
g}(\alpha_{s},x)=\sum_{n=1}(\frac{\alpha_{s}}{4\pi})^{n}c^{(n)}_{L,q
\& g}(x)$ [30,31,32] where $n$ denotes the order of analysis at
leading order (i.e., $n=1$) up to high order analysis (i.e.,
$n>1$). All further theoretical details relevant for analyzing
$F_{L}$ at NLO and NNLO in the $\overline{\mathrm{MS}}$
factorization scheme have been presented in [30,31,32]. In GM-VFNS
the transition from $n_{f}$ active flavors to $n_{f}+1$ considered
in the construction of charm-quark parton distribution function.
Rather at some large scales the transition with two massive quarks
(i.e., $n_{f}{\rightarrow}n_{f}+2$) has been discussed in
Refs.[33-38]. For $Q^{2}$ below the $c$-quark threshold
($Q^{2}<m^{2}_{c}$), $n_{f}$ in (1) should be replaced by
$n_{f}-1$ and for $Q^{2}$ above the $b$-quark threshold $n_{f}$
should be replaced by $n_{f}+1$. For $n_{f}=4$ where
$<e^{2}>=\frac{5}{18}$, Eq.(1) reads as
\begin{eqnarray}
F_{L}(x,Q^{2})=C_{L,q}\otimes
F_{2}(x,Q^{2})+\frac{5}{18}C_{L,g}\otimes G_{n_{f}=4}(x,Q^{2}).
\end{eqnarray}
On the threshold of heavy quark production, the number of active
flavours $n_{f}$ is 3 and the heavy flavour cross section is
generated by photon-gluon fusion (PGF). Now I turn to the
perturbative predictions for longitudinal structure function in
accordance with the heavy quark production threshold which can be
written as [39]
\begin{eqnarray}
F_{L}(x,Q^{2})&=&C_{L,q}\otimes
F_{2}(x,Q^{2})+\frac{2}{9}C_{L,g}\otimes
G_{n_{f}=3}(x,Q^{2})\nonumber\\
&&+F_{L}^{c}(x,Q^{2}),~\mathrm{for}~Q^{2}{\geq}m_{c}^{2}
\end{eqnarray}
and
\begin{eqnarray}
 F_{L}(x,Q^{2})&=&C_{L,q}\otimes
F_{2}(x,Q^{2})+\frac{2}{9}C_{L,g}\otimes
G_{n_{f}=3}(x,Q^{2})\nonumber\\
&&+F_{L}^{c}(x,Q^{2})+F_{L}^{b}(x,Q^{2}),~\mathrm{for}~Q^{2}{\geq}m_{b}^{2}.
\end{eqnarray}
In FFNS heavy quarks are not considered as active. In this case,
for $Q^{2}{\sim}m^{2}_{\mathcal{Q}}$ ($\mathcal{Q}=c,b$) the heavy
flavors are generated only by PGF. In the FFNS at low $Q^{2}$, the
longitudinal heavy-quark structure function at low values of $x$
is given by
\begin{eqnarray}
F_{L}^{\mathcal{Q}\overline{\mathcal{Q}}}(x,Q^{2})=C_{L,g}^{\mathcal{Q}\overline{\mathcal{Q}}}(x,\xi){\otimes}
G_{n_{f}}(x,\mu^{2}),
\end{eqnarray}
where $n_{f}=3$ is the number of light quark flavors when all
heavy flavors are considered as massive, and $G_{n_{f}}$ is the
gluon distribution function due to the number of active quark
flavors. In the GM-VFNS at high $Q^{2}$, the heavy-flavor
structure functions are dependence to the active flavor number as
I take $n_{f}=4$ for $m_{c}^{2}<\mu^{2}<m_{b}^{2}$ and $n_{f}=5$
for $m_{b}^{2}<\mu^{2}<m_{t}^{2}$. Within this scheme, heavy quark
densities arise via the
$g{\rightarrow}\mathcal{Q}\overline{\mathcal{Q}}$ evolution. In
the small-$x$ range the heavy quark contributions are given by
these forms:
\begin{eqnarray}
F_{L}^{c}(x,Q^{2})=C_{L,g}^{c\overline{c}}(x,\xi){\otimes}G_{n_{f}=4}(x,\mu^{2}),\nonumber\\
F_{L}^{b}(x,Q^{2})=C_{L,g}^{b\overline{b}}(x,\xi){\otimes}G_{n_{f}=5}(x,\mu^{2}).
\end{eqnarray}
where $\xi=\frac{m^{2}_{c,b}}{Q^{2}}$. The heavy flavor
contribution to $F_{L}$ is taken as given by fixed-order NLO
perturbative theory. The coefficient functions up to NLO
approximation were demonstrated in Refs.[40-44] as
\begin{eqnarray}
C_{L,g}(z,\xi)=C^{0}_{L,g}(z,\xi)+\frac{\alpha_{s}(\mu^{2})}{4\pi}C^{1}_{L,g}(z,\xi).
\end{eqnarray}
The default renormalisation scale $\mu_{r}$ and factorization
scale $\mu_{f}$ are set to $\mu\equiv
\mu_{r}=\mu_{f}=\sqrt{Q^{2}+4m^2_{c,b}}$. In this scheme on should
take quark mass into account,  as  the rescaling variable $\chi$
is defined into the Bjorken variable $x$ by the following form
[45]
\begin{eqnarray}
\chi=x(1+\frac{4m_{\mathcal{Q}}^{2}}{Q^{2}}),
 \end{eqnarray}
where the rescaling variable, at high $Q^{2}$ values
($m^{2}_{c,b}/Q^{2}\ll 1$), reduces to $x$ as
$\chi{\rightarrow}x$. Recently several methods for the
determination of the heavy quark longitudinal structure function
in the nucleon  have been proposed [46-59]. Recently, the ratio of
structure functions in the heavy quark production processes,
obtained using the the transverse momentum dependent  gluon
distribution function, can be found in [60].\\
 The parameterization
of $F_{2}(x,Q^{2})$ and $G_{n_{f}}(x,Q^{2})$ have been suggested
by authors in Refs.[11] and [12] respectively. These
parameterizations obtained from a combined fit of HERA data [61]
for $x<0.1$ and over a wide range of $Q^{2}$ values. The proton
structure function parameterized with a  global fit function [12]
to the ZEUS data for $F_{ 2}(x,Q^{2})$ for $0.11< Q^{2} < 1200~
GeV^{2}$ and $x< 0.1$ takes the form
\begin{eqnarray}
F_{ 2} (x,Q^{2})& =& (1- x)[\frac{F_{P}}{1 -x_{P}} +
A(Q^{2})\ln( \frac{x_{P}}{ x}\frac{1 - x}{ 1 - x_{P}})\nonumber\\
&&+B(Q^{2}) \ln^{2}( \frac{x_{P}}{ x}\frac{1 - x}{ 1 - x_{P}})],
\end{eqnarray}
where
\begin{eqnarray}
 A(Q^{2}) = a_{0} + a_{1} {\ln}Q^{2} + a_{2} {\ln}^{2}
 Q^{2},\nonumber
 \end{eqnarray}
and
\begin{eqnarray}
  B(Q^{2}) = b_{0} + b_{1}
{\ln}Q^{2} + b_{2} {\ln}^{2} Q^{2}.\nonumber
\end{eqnarray}
The fitted parameters are tabulated in Table I. In the case of
four massless quarks the gluon distribution function is obtained
with an expression quadratic in both $\ln{Q^{2}}$ and $\ln{(1/x)}$
for $0<x ~{\leq}~ 0.06$ as
\begin{eqnarray}
G_{nf=4}(x,Q^{2})& =& \frac{3}{5}G_{nf=3}(x,Q^{2}),
\end{eqnarray}
where
\begin{eqnarray}
G_{nf=3}(x,Q^{2})&=&-2.94-0.359~{\ln}Q^{2}-0.101~{\ln}^{2}Q^{2}\nonumber\\
&&+(0.594-0.0792~{\ln}Q^{2}-0.000578~{\ln}^{2}Q^{2})\nonumber\\
&&{\times}\ln(1/x)+(0.168+0.138~{\ln}Q^{2}\nonumber\\
&&+0.0169~{\ln}^{2}Q^{2})\ln^{2}(1/x),
\end{eqnarray}
and
\begin{eqnarray}
G_{nf=5}(x,Q^{2})& =& \frac{6}{11}G_{nf=3}(x,Q^{2}).
\end{eqnarray}
Coefficients and more discussions about them can be seen in
Refs.[11] and [12]. The massless distributions for $n_{f}=4$ and
$n_{f}=5$ massless quarks are just $3/5$ and $6/11$ of
$G_{nf=3}(x,Q^{2})$ [12]. In Ref.[15] the QCD parameter $\Lambda$
has been extracted due to $\alpha_{s}(M_{z}^{2})=0.1166$, which
corresponds to the number of active flavor at NLO approximation as
$ \Lambda(n_{f}=5)=195.7~ \mathrm{MeV}, \Lambda(n_{f}=4)=284.0~
\mathrm{MeV}$ and $\Lambda(n_{f}=3)=347.2~ \mathrm{MeV}$.\\
With the explicit form of the distribution functions, I can
proceed to extract the longitudinal structure function
$F_{L}(x,Q^{2})$ from data mediated by the parametrization of $F_{
2}(x,Q^{2})$ and $G_{n_{f}}(x,Q^{2})$. I have calculated the
$x$-dependence of the longitudinal structure function at several
fixed values of $Q^{2}$ corresponding to H1-Collaboration data
[1,2]. Results are presented and compared with H1 data [1,2] and
the parameterization of $F_{L}(x,Q^{2})$ [15] at NLO approximation
in Fig.1. It is seen that, for all values of the presented
$Q^{2}$, the extracted longitudinal structure function within the
NLO approximation at $n_{f}=4$ is in a much better agreement with
data and is comparable with the Mellin transforms method
[13,14,15]. Also the impact of the gluon distribution due to the
number of active flavor in the heavy-quark longitudinal structure
function is considered. The running charm and bottom-quak masses
are considered as $m_{c}=1.29~\mathrm{GeV}$ and
$m_{b}=4.049~\mathrm{GeV}$ [62]. In order to present more detailed
discussions on our findings,  the results for the longitudinal
structure function compared with CT18 [63] in this figure. As can
be seen from the related figures, the longitudinal structure
function results are consistent with the CT18 NLO at moderate and
large values of $Q^{2}$.\\
In Fig.2, I have calculated the $Q^{2}$-dependence of the
longitudinal structure function at low $x$ in the NLO
approximation. Results of calculations and comparison with data of
the H1-Collaboration [1,2] are presented in this figure (i.e.,
Fig.2), where the dashed-dot (Ref.[15]) and dashed-dot-dot (this
work) lines correspond to the extracted $F_{L}(x,Q^{2})$ at
$n_{f}=4$ in the NLO approximation, respectively. These results
have been performed at fixed value of the invariant mass $W$ as
$W=230~ \mathrm{GeV}$. These results are compared with the CT18
NLO. Figure 2 shows that the parameterization of parton
distributions provides correct behaviors of the extracted
$F_{L}(x,Q^{2})$ within the NLO approximation. Over a wide range
of variable $Q^{2}$, the extracted longitudinal structure
functions are in a good agreement with experimental data in
comparison with the parameterization of $F_{L}(x,Q^{2})$. At low
values of $Q^{2}$, the extracted results are  still above the
experimental data. Although this result is consistent with the
Mellin transforms method in this region.\\
In order to make sure these results are in the deep inelastic
region, I use the nonlinear longitudinal structure function
 where effects of the nonlinear corrections to the gluon distribution
are taken into account. The most important correction to the gluon
distribution is the gluon recombination effect. For small momentum
transfer the produced gluon overlap themselves in the transverse
area and fusion processes, $gg{\rightarrow}g$, become important
[64-70]. This gluon recombination effect is one of the important
correction to the DGLAP equations [71,72,73] which theoretical
predictions of this effect is initialled by Gribov, Levin and
Ryskin [74] and followed by Mueller and Qiu (MQ) [75].\\
The GLR-MQ equation can be written in standard form [76,77,78]
\begin{eqnarray}
\frac{\partial{G(x,Q^{2})}}{\partial{\ln}Q^{2}}&=&\frac{\partial{G(x,Q^{2})}}{\partial{\ln}Q^{2}}|_{DGLAP}\nonumber\\
&&-\frac{81}{16}\frac{\alpha^{2}_{s}(Q^{2})}{\mathcal{R}^{2}Q^{2}}\int_{\chi}^{1}\frac{dz}{z}G^{2}(\frac{x}{z},Q^{2}),
\end{eqnarray}
where $\chi=\frac{x}{x_{0}}$ and $x_{0}$ is the boundary condition
that the gluon distribution joints smoothly onto the linear
region. The correlation length $\mathcal{R}$  determines the size
of the nonlinear terms. This value depends on how the gluon
ladders are coupled to the nucleon or on how the gluons are
distributed within the nucleon. By solving GLR-MQ (i.e., Eq.13),
the nonlinear correction to  the gluon distribution function
(i.e., $G_{n_{f}}^{\mathrm{NL}}(x,Q^{2})$ ) is obtained by the
following form as
\begin{eqnarray}
G_{n_{f}}^{\mathrm{NL}}(x,Q^{2})&=&G_{n_{f}}^{\mathrm{NL}}(x,Q_{0}^{2})+G_{n_{f}}(x,Q^{2})-G_{n_{f}}(x,Q_{0}^{2})\nonumber\\
&&-\int_{Q_{0}^{2}}^{Q^{2}}\frac{81}{16}\frac{\alpha^{2}_{s}(Q^{2})}{\mathcal{R}^{2}Q^{2}}\int_{\chi}^{1}\frac{dz}{z}G_{n_{f}}^{2}(\frac{x}{z},Q^{2})d{\ln}Q^{2}\nonumber\\
\end{eqnarray}
At $Q_{0}^{2}$ the low $x$ behavior of the nonlinear gluon
distribution is assumed to be [79,80,81]
\begin{eqnarray}
G_{n_{f}}^{\mathrm{NL}}(x,Q_{0}^{2})&=&G_{n_{f}}(x,Q_{0}^{2})\{1+\frac{27\pi{\alpha_{s}(Q_{0}^{2})}}{16\mathcal{R}^{2}Q_{0}^{2}}\theta(x_{0}-x)\nonumber\\
&&{\times}[G_{n_{f}}(x,Q_{0}^{2})-G_{n_{f}}(x_{0},Q_{0}^{2})]
\}^{-1}.
\end{eqnarray}
In Fig.3, AM equation with GLR-MQ correction is used to evaluate
the longitudinal structure function at low $x$ and $Q^{2}$. As can
be seen in Fig.3, the nonlinear correction is very important to
slow down the longitudinal structure function behavior at low
$Q^{2}$ values. The evolutions of the nonlinear correction to
$F_{L}$ with $Q^{2}$ at fixed value of the invariant mass $W$ and
the comparisons with the experimental data [1] and CT18 [63] are
shown in Fig.3. A comparison between Figs.2 and 3 shows that the
nonlinear effects of the longitudinal structure function are
observable for $x<x_{0}=0.01$ at hotspot point where gluons are
populated across the proton as it is equal to $\mathcal{R} \simeq
2~\mathrm{GeV}^{-1}$. As can be seen, the nonlinear results at hot
spot point at low and moderate $Q^{2}$ values seem to be
compatible with the H1 data and CT18 at NLO and NNLO
approximations. Indeed he nonlinear corrections here are negative
and result in a better agreement with data and
 parameterization method.\\
To summarize, I used the parton parameterization method suggested
in Refs.[11,12] to obtain the longitudinal structure function. The
method depends on  the number of  active flavor as  the results
are considered in two cases $n_{f}=4$ and $n_{f}=3+c+b$. The heavy
flavor contributions to the longitudinal structure function are
taken as $Q^{2}$ increases above about $m^{2}_{c}$ and then above
about $m^{2}_{b}$. Dependent gluon distributions are usually are
given in a parameterization method in which the number of active
quark flavors increases from $n_{f}=3$ to $n_{f}=4$ and then to
$n_{f}=5$ corresponds to the threshold heavy quark production,
which are directly related to the gluon density in the photo-gluon
fusion reactions. I have applied the nonlinear correction to
extract the longitudinal structure function at low values of $x$.
It has been found that at low and moderated $Q^{2}$, NLO results
are corresponding to the experimental data and other
parameterization methods.
\subsection{ACKNOWLEDGMENTS}
The author is thankful to Razi University for financial support of
this project. The author is especially grateful to A.V.Kotikov for
carefully reading the manuscript and for critical notes. Thanks
also go to H.Khanpour for help with preparation of the
QCD parametrization model.\\
\begin{table}[h]
\caption{ Parameters of Eq.(8) [12], resulting from a global fit
to the ZEUS data.}
\begin{tabular} {cccc}
\toprule \\  \multicolumn{2}{c}{parameters \quad \quad \quad ~~~~~~~~~~~~~~~~value}    \\ &&&\\ \hline \\ &&&\\
  $a_{0} $  &   \quad  $-7.828\times 10^{-2}\pm5.19\times10^{-3} $  \\
  $a_{1} $  &   \quad   $2.248\times 10^{-2}\quad\pm1.47\times10^{-3}$  \\
  $a_{2}$   &    \quad  $2.301\times 10^{-4}\pm4.88\times10^{-4}$   \\  &&&\\
 $b_{0}$   &   \quad   $1.313\times 10^{-2}\pm6.99\times10^{-4} $ \\

 $b_{1}$   &   \quad   $4.736\times 10^{-3}\pm2.98\times10^{-4}$  \\

 $b_{2}$    &    \quad  $1.064\times 10^{-3}\pm3.88\times10^{-5} $ \\ &&& \\
$F_{p}$& \quad  $0.503\pm 0.0012$ & &\\
$x_{p}$& \quad  $0.0494\pm 0.0039$ & &\\
$\chi^{2}(\mathrm{goodness~ of~ fit})$ &  \quad  $1.11$ & &\\
\hline

\end{tabular}
\end{table}
\newpage{
\subsection{References}
1. H1 Collab. (V.Andreev , A.Baghdasaryan, S.Baghdasaryan et al.),
Eur.Phys.J.C{\bf74},
2814(2014).\\
2. H1 Collab. (F.D.Aaron , C.Alexa, V.Andreev et al.),Eur.Phys.J.C{\bf71}, 1579 (2011).\\
3. ZEUS Collab. (H.Abramowicz , I.Abt, L.Adamczyk et al.), Phys.Rev.D{\bf90}, 072002 (2014).\\
4. P.Agostini , H.Aksakal, H.Alan et al. [LHeC Collaboration and FCC-he Study Group ], CERN-ACC-Note-2020-0002, arXiv:2007.14491 [hep-ex] (2020).\\
5. J.Bartels, K.Golec-Biernat and  L.Motyka, Phys.Rev. D{\bf81}, 054017 (2010).\\
6. V.Tvaskis , A.Tvaskis, I.Niculescu et al., Phys.Rev.C{\bf97}, 045204(2018).\\
7. V.Chekelian, arXiv [hep-ex]:0810.5112 (2008).\\
8. G.Altarelli and G. Martinelli, Phys.Lett.B{\bf 76}, 89(1978).\\
9. R.S.Thorne, arXiv:hep-ph/9805298(1998).\\
10. A.D.Martin W.J.Stirling and R.S.Thorne, Phys.Lett.B{\bf 636}, 259(2006).\\
11. M.M.Block, L.Durand and D.W.McKay, Phys.Rev.D{\bf77}, 094003(2008).\\
12. M.M.Block and L.Durand, arXiv: 0902.0372 [hep-ph](2009).\\
13. L.P.Kaptari , A.V.Kotikov, N.Yu.Chernikova and P.Zhang, JETP Lett.{\bf 109}, 281(2019).\\
14. A.V.Kotikov, JETP Lett.{\bf111}, 67 (2020).\\
15. L.P.Kaptari , A.V.Kotikov, N.Yu.Chernikova and P.Zhang, Phys.Rev.D{\bf 99}, 096019(2019).\\
16. B. Rezaei and G.R. Boroun, Eur.Phys.J.A{\bf56}, 262 (2020).\\
17. G.R. Boroun, Phys.Rev.C{\bf97}, 015206 (2018).\\
18. G.R. Boroun, Eur.Phys.J.Plus{\bf129}, 19 (2014).\\
19. G.R. Boroun and B. Rezaei, Eur.Phys.J.C{\bf72}, 2221
(2012).\\
20. N. Baruah, M.K. Das and J.K. Sarma,  Eur.Phys.J.Plus{\bf129},
229 (2014).\\
21. G.R.Boroun and B.Rezaei, EPL,{\bf133}, 61002 (2021).\\
22. G.R.Boroun, Chin.Phys.C{\bf45}, 063105 (2021).\\
23. G.R.Boroun and B.Rezaei, Phys.Lett.B{\bf816}, 136274 (2021).\\
24. G.R.Boroun, Eur.Phys.J.Plus{\bf135}, 68(2020).\\
25. G.R.Boroun and B.Rezaei, Nucl.Phys.A{\bf1006}, 122062(2021).\\
26. B.Rezaei and G.R.Boroun, Commun.Theor.Phys.{\bf59}, 462
(2013).\\
27. G.R.Boroun and B.Rezaei, Eur.Phys.J.C{\bf72}, 2221 (2012).\\
28. B.Rezaei and G.R.Boroun, Nucl.Phys.A{\bf857}, 42(2011).\\
29. G.R.Boroun, Int.J.Mod.Phys.E{\bf18}, 131(2009).\\
30. S.Moch, J.A.M.Vermaseren and A.Vogt, Phys.Lett.B{\bf 606},
123(2005).\\
31. S.Alekhin, J.Bl$\ddot{\mathrm{u}}$mlein and S.-O.Moch, arXiv[hep-ph]:1808.08404 (2018).\\
32. A.M.Cooper-Sarkar, R.C.E.Devenish and A.De Roeck, Int.J.Mod.Phys.A{\bf13}, 3385 (1998).\\
33. R.Thorne, Phys.Rev.D{\bf73}, 054019 (2006).\\
34. R.Thorne, Phys.Rev.D{\bf86}, 074017 (2012).\\
35. S.Alekhin, J. Bl$\ddot{\mathrm{u}}$mlein, S. Klein and S. Moch , arXiv [hep-ph]:0908.3128(2009).\\
36. G.Beuf, C.Royon and D.Salek, arXiv[hep-ph]:0810.5082(2008).\\
37. J.Bl$\ddot{\mathrm{u}}$mlein, A.De Freitas, C.Schneider and
K.Sch$\ddot{\mathrm{o}}$nwald, Phys. Lett.B{\bf782},
 362(2018).\\
38.  S.Alekhin, J. Bl$\ddot{\mathrm{u}}$mlein and S. Moch, arXiv [hep-ph]:2006.07032(2020).\\
39. M.Gl$\mathrm{\ddot{u}}$ck, C.Pisano and E.Reya,
Phys.Rev.D\textbf{77},
074002(2008).\\
40. S.Riemersma, J.Smith and W.L.van Neerven, Phys.Lett.B{\bf 347}, 143(1995).\\
41.  A.V.Kisselev, Phys.Rev.D{60}, 074001(1999).\\
42.  E.Laenen, S.Riemersma, J.Smith and W.L. van Neerven,
Nucl.Phys.B\textbf{392}, 162(1993).\\
43. A.Y.Illarionov, B.A.Kniehl and A.V.Kotikov, Phys.Lett.B{\bf 663}, 66 (2008).\\
44. S.Catani and F.Hautmann, Nucl.
Phys.B\textbf{427}, 475(1994).\\
45. Wu-Ki Tung, S. Kretzer, and C. Schmidt, J. Phys. G {\bf28},
983 (2002).\\
46. G.R. Boroun, B. Rezaei and J.K. Sarma, Int.J.Mod.Phys.A{\bf29}
(2014) 32, 1450189.\\
47. G.R. Boroun and B. Rezaei, Nucl.Phys.A{\bf929}, 119(2014).\\
48. G.R. Boroun,  Nucl.Phys.B{\bf884}, 684(2014).\\
49. G.R.Boroun, PoSHQL2012(2012)069.\\
50. G.R. Boroun and B. Rezaei, EPL{\bf100}, 41001(2012).\\
51. G.R. Boroun and B. Rezaei, Nucl.Phys.B{\bf857}, 143(2012).\\
52. G.R. Boroun and B. Rezaei, J.Exp.Theor.Phys.{\bf115}, 427(2012).\\
53.  Ali N. Khorramian, S. Atashbar Tehrani and A. Mirjalili,
Nucl.Phys.B Proc.Suppl.{\bf186}, 379 (2009).\\
54. S.Khatibi and H.Khanpour, Nucl.Phys.B{\bf967}, 115432 (2021).\\
55. J.Blumlein, A.De Freitas, W.L. van Neerven and S.Klein, Nucl.Phys.B{\bf755}, 272 (2006).\\
56. H.Khanpour, Nucl.Phys.B{\bf958}, 115141 (2020).\\
57. H.Khanpour,Phys.Rev.D{\bf99}, 054007 (2019).\\
58. G.R.Boroun,
Phys.Lett.B{\bf744}, 142 (2015).\\
59. G.R.Boroun, Phys.Lett.B{\bf741}, 197 (2015).\\
60. A.V.Kotikov, A.V.Lipatov and P.Zhang, arXiv
[hep-ph]:2104.13462
(2021).\\
61. H1 and ZEUS Collab. (F.D.Aaron, H.Abramowicz, I.Abt et al.),
JHEP {\bf1001}, 109
(2010).\\
62. H1 and ZEUS Collab. (H.Abramowicz, I.Abt, L.Adamczyk  et al.),
Eur.Phys.J.C{\bf78}, 473(2018).\\
63. Tie-Jiun Hou et al., Phys.Rev.D{\bf103}, 014013(2021).\\
64. A.V.Giannini and F.O.Dur$\widetilde{a}$es, Phys.Rev.D{\bf88}, 114004(2013).\\
65. G.R.Boroun and B.Rezaei, arXiv [hep-ph]:2105.01121 (2021).\\
66. R.Wang and X.Chen, Chinese Phys.C{\bf41}, 053103(2017).\\
67. G.R.Boroun and S.Zarrin, Eur.Phys.J.Plus{\bf128}, 119(2013).\\
68. B.Rezaei and G.R.Boroun, Phys.Lett.B{\bf692}, 247(2010).\\
69. G.R.Boroun, Eur.Phys.J.A{\bf43}, 335(2010).\\
70. G.R.Boroun, Eur.Phys.J.A{\bf42}, 251(2009).\\
71. Y. L. Dokshitzer, Sov. Phys. JETP{\bf 46}, 641 (1977).\\
72. V. N. Gribov and L. N. Lipatov, Sov. J. Nucl. Phys.{\bf15},
438 (1972).\\
73. G. Altarelli and G. Parisi, Nucl. Phys. B {\bf126}, 298 (1977).\\
74. L. V. Gribov, E. M. Levin and M. G. Ryskin, Phys. Rep.
{\bf100}, 1 (1983).\\
75. A. H. Mueller and Jianwei Qiu, Nucl. Phys. B {\bf268}, 427
(1986).\\
76. K.Prytz, Eur.Phys.J.C{\bf22}, 317(2001).\\
77. K.J.Eskola,H.Honkanen, V.J.Kolhinen, J.Qiu and C.A.Salgado, Nucl.Phys.B{\bf660}, 211(2003).\\
78. M.A Kimber, J.Kwiecinski and
A.D.Martin, Phys.Lett.B{\bf508}, 58(2001).\\
79. A.D.Martin, W.J Stirling and R.G Roberts, Phys.Rev.D{\bf47}, 867(1993).\\
80. J.Kwiecinski, A.D.Martin and P.J.Sutton, Phys.Rev.D{\bf44},
2640(1991).\\
81. A.J.Askew, J.Kwiecinski,
A.D.Martin and P.J.Sutton, Phys.Rev.D{\bf47}, 3775(1993).\\
}
\newpage
\begin{figure} \centering
\includegraphics[width=1\textwidth]{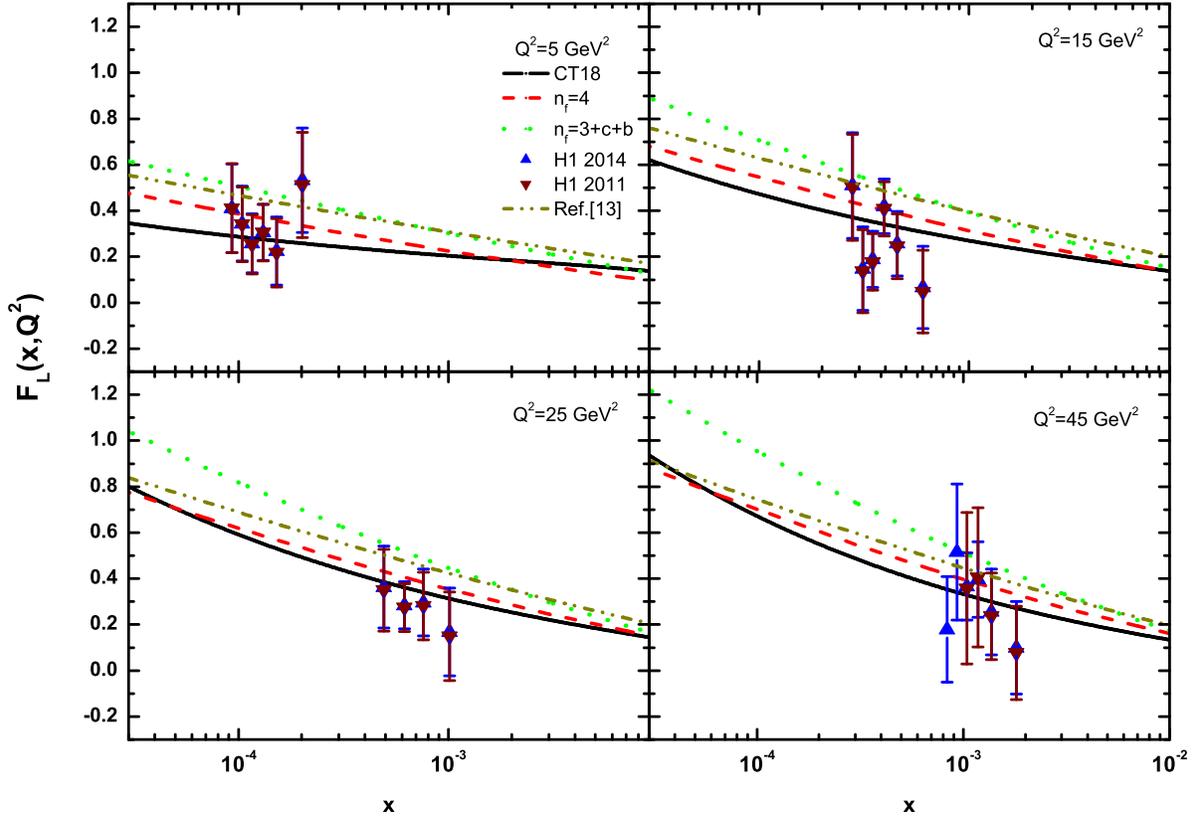}
\caption{The longitudinal structure function $F_{L}(x,Q^{2})$
extracted from the parameterization of parton distributions at
fixed $Q^{2}$ as a function of $x$ variable. The solid and
dashed-dot-dot lines represent the CT18 [63] and the Mellin
transforms method [15] at the NLO approximation respectively. The
dashed and dot lines represent our predictions at $n_{f}=4$ and
$n_{f}=3+c+b$ at the NLO approximation respectively. Experimental
data (up-triangle H1 2014, down-triangle H1 2011 ) are from the
H1-Collaboration [1,2] as accompanied with total errors.}
\label{Fig1}
\end{figure}
\begin{figure} \centering
\includegraphics[width=1\textwidth]{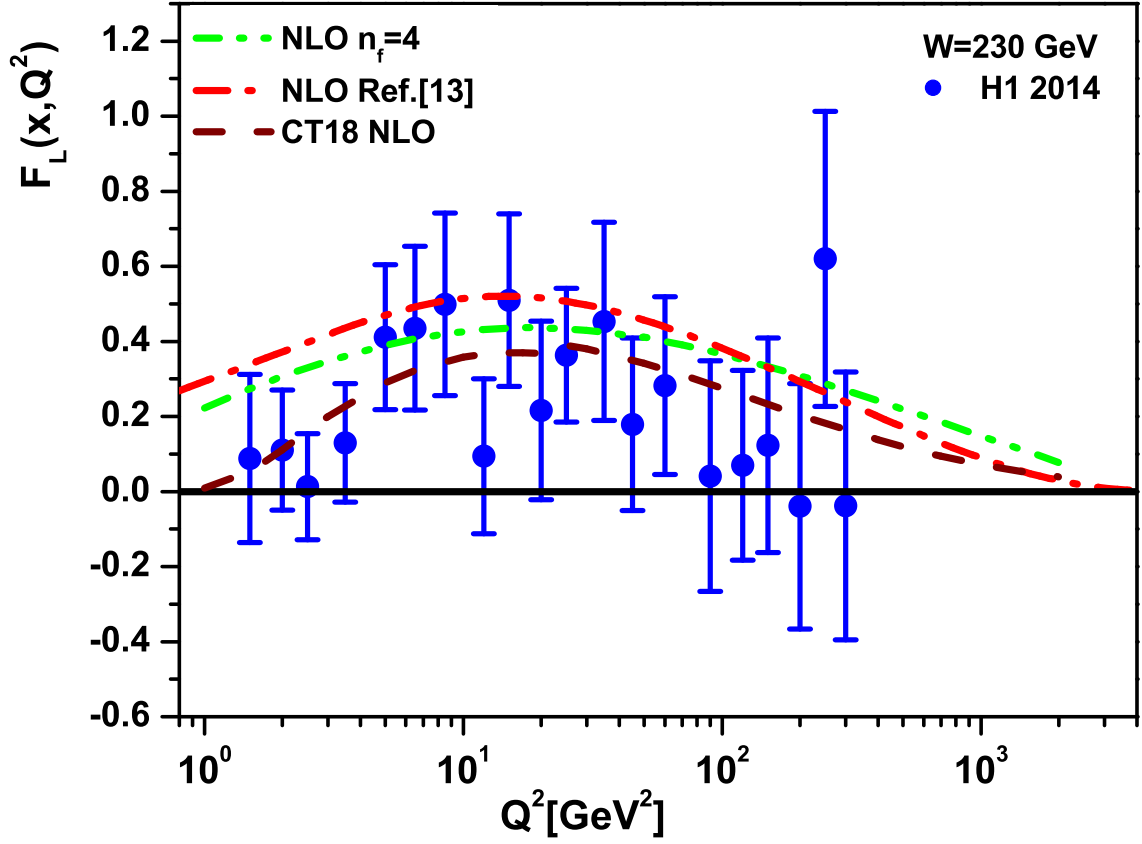}
\caption{The obtained longitudinal structure function
$F_{L}(x,Q^{2})$ from the parameterization of parton distributions
as a function of variable $Q^{2}$ at fixed value of the invariant
mass $W=230~\mathrm{GeV}$. The dashed and dashed-dot lines
represent the CT18 [63] and the Mellin transforms method [15] at
the NLO approximation. The dashed-dot-dot line represents our
predictions at $n_{f}=4$ within the NLO approximation.
Experimental data are from the H1-Collaboration [1] as accompanied
with total errors. } \label{Fig1}
\end{figure}
\begin{figure} \centering
\includegraphics[width=1\textwidth]{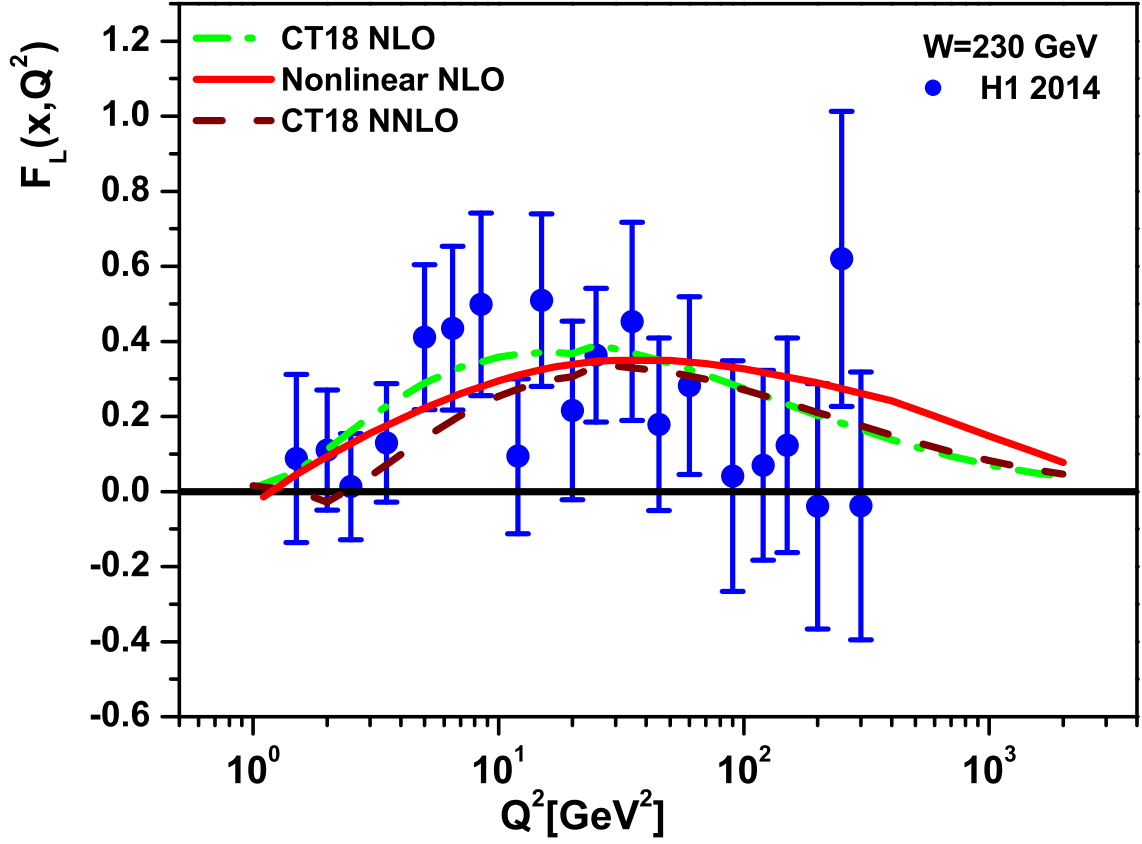}
\caption{The obtained longitudinal structure function
$F_{L}(x,Q^{2})$ from the nonlinear gluon distribution as a
function of variable $Q^{2}$ at fixed value of the invariant mass
$W=230~\mathrm{GeV}$ at hotspot point. The dashed and dashed-dot
lines represent the CT18 [63] at the NNLO and NLO approximation
respectively. The solid line represents nonlinear behavior of the
longitudinal structure function  at $n_{f}=4$ within the NLO
approximation. Experimental data are from the H1-Collaboration [1]
as accompanied with total errors. } \label{Fig1}
\end{figure}
\end{document}